\documentclass{iaus}
\usepackage{graphicx}

\newcommand{\ebv}{$E_{B-V}$}

\newcommand{\rv}{$R(V)$}

\newcommand{\foiii}{[O~{\sc iii}]}

\title[Mapping the 3D extinction and DIBs with LAMOST] 
{Mapping the three-dimensional multi-band extinction and diffuse interstellar bands in the Milky Way with LAMOST}

\author[H.-B. Yuan et al.]   
{H.-B. Yuan$^{1,2}$, X.-W. Liu$^{1,3}$, M.-S Xiang$^3$, Z.-Y. Huo$^4$, H.-H. Zhang$^3$, Y. Huang$^3$
\and H.-W. Zhang$^3$} 
\affiliation{$^1$ Kavli Institute for Astronomy and Astrophysics, Peking University, Beijing 100871, China\\ 
email: {\tt yuanhb4861@pku.edu.cn} \\[\affilskip]
$^2$ LAMOST Fellow  \\ 
$^3$ Department of Astronomy, Peking University, Beijing 100871, China \\
$^4$ National Astronomical Observatories, Chinese Academy of Sciences, Beijing 100012, China 
}

\pubyear{2014}
\volume{298}  
\pagerange{XX -- YY}
\jname{IAUS\,298 Setting the scence for Gaia and LAMOST}
\editors{S. Feltzing, G. Zhao, N.\,A. Walton \& P.\,A. Whitelock, eds.}
\begin{document}

\maketitle

\begin{abstract}
With modern large scale spectroscopic surveys, such as the SDSS and
LSS-GAC, Galactic astronomy has entered the era of millions of stellar spectra.
Taking advantage of the huge spectroscopic database, we propose
to use a "standard pair" technique to a) Estimate multi-band extinction towards sightlines of millions of
stars; b) Detect and measure the diffuse interstellar bands in hundreds of thousands SDSS and LAMOST 
low-resolution spectra; c) Search for extremely faint emission line nebulae in the Galaxy; and d) 
Perform photometric calibration for wide field imaging surveys.
In this contribution, we present some results of applying this technique to the SDSS data,
and report preliminary results from the LAMOST data.
\keywords{ISM: dust, extinction, ISM: lines and bands, planetary nebulae: general, ISM: general, surveys,
techniques: spectroscopic}
\end{abstract}

\firstsection 
\section{Overview}
Dust grains produce extinction and reddening of stellar light from
the ultraviolet (UV) to the infrared (IR) (Draine 2003).
Accurate determination of reddening to a star is vital for reliable derivation of
its basic stellar parameters, such as effective temperature and distance.
Constructing a 3D Galactic extinction map
plays an essential role in Galactic astronomy,
particularly in achieving the driving goals of the LAMOST
Spectroscopic Survey of the Galactic Anti-center (LSS-GAC; Liu et al. this volume).

The Sloan Digital Sky Survey (SDSS; York et al. 2000) has delivered low-resolution spectra for
about 0.7\,M stars in its Data Release 9 (DR9; Ahn et al. 2012).  
The LAMOST Galactic surveys (Deng et al. 2012 and this volume) 
have obtained over 1\,M stellar spectra and will obtain over 5\,M in the next four years. 
With millions of stellar spectra, "identical" stars in different environments can be easily paired and compared,
which opens great opportunities to a number of studies with 
the standard pair technique (Stecher 1965; Massa, Savage \& Fitzpatrick 1983). 
By comparing the differences in photometric colors from large scale imaging surveys, 
such as the Galaxy Evolution Explore (GALEX; Martin et al. 2005) in the UV, the SDSS in the optical, 
the Two Micro All Sky Survey (2MASS; Skrutskie et al. 2006) in the near-IR and the Wide-field Infrared Survey Explorer 
(WISE; Wright et al. 2010) in the mid-IR, 
one can measure multi-band reddening for a large number of 
targets and constrain the reddening laws (Yuan, Liu \& Xiang 2013).
By comparing the differences in normalized spectra, one can detect absorption features from the 
interstellar medium (ISM), particularly the diffuse interstellar bands (DIBs; Yuan \& Liu 2012),
as well as abnormal stellar absorption/emission lines from chemically particular or active stars.
Such method has the advantages that it's straight-forward, model-free and applicable to the majority of stars.
Combining stellar distances from Gaia (Perryman et al. 2001) or from spectro-photometry, one can further 
map the Galactic extinction, extinction laws and DIBs in 3D.
If by chance a diffuse nebula falls in the sightline of some targets, 
lines emitted by the nebula (e.g. \foiii~$\lambda\lambda$4959, 5007) will also be recorded and
can be used to discover new faint emission line nebulae (e.g. planetary nebulae and supernova remnants) 
in the Galaxy (Yuan \& Liu, submitted). 
Finally, the technique can also be used to perform photometric calibration for wide field imaging surveys (Yuan et al. in prep).
In the following sections, we use the cases above to demonstrate the power of the 
standard pair technique when applied to  large scale spectroscopic and photometric datasets including
the SDSS and on-going LAMOST surveys.

\section{The 3D multi-band extinction and extinction laws}

Using star pairs selected from the SDSS, and combining the SDSS, GALEX, 2MASS and WISE  
photometry ranging from the far-UV to the mid-IR, 
Yuan, Liu \& Xiang (2013) have measured dust reddening in the $FUV-NUV, NUV-u, u-g,
g-r, r-i, i-z, z-J, J-H, H-Ks, Ks-W1$ and $W1-W2$ colors for thousands of
Galactic stars.  The measurements, together with the \ebv~values given by Schlegel et al. (1998),
allow us to derive the empirical reddening coefficients for those colors.
The results are compared with previous measurements and the predictions of a variety of
Galactic reddening laws. We find that 1) The dust reddening map of Schlegel et
al. (1998) over-estimates \ebv~by about 14\%, consistent with the work of
Schlafly et al. (2010) and Schlafly \& Finkbeiner (2011); 
and 2) All the new reddening coefficients, except those for $NUV-u$ and $u-g$,  prefer the \rv~=~3.1 Fitzpatrick
reddening law (Fitzpatrick 1999) rather than the \rv~=~3.1 CCM (Cardelli et al. 1989) and 
O'Donnell (O'Donnell 1994) reddening laws. Using the $Ks$-band extinction
coefficient predicted by the \rv~=~3.1 Fitzpatrick law and the
observed reddening coefficients, we have deduced new extinction
coefficients for the $FUV, NUV, u, g, r, i, z, J, H, W1$ and $W2$ passbands.
We recommend that the new reddening and extinction coefficients should be used in
the future and an update of the Fitzpatrick reddening law in the UV is probably necessary.

LSS-GAC has obtained about 0.8\,M spectra of ${\rm
S/N}(\lambda7450) \ge 10$ per pixel and basic stellar parameters 
for about 0.6\,M spectra of ${\rm S/N}(\lambda4650) > 10$ per pixel (Liu et al. this volume). 
With the same technique, we measured dust reddening in the $g-r, r-i, i-z, z-J, J-H$ and  $H-Ks$
colors for over 0.2\,M stars from LSS-GAC.
The reddening coefficients for these colors relative to that for $g-r$ are 
consistent with the work of Yuan, Liu \& Xiang (2013), as seen in Fig.\,1.
With spectro-photometric distances, we have constructed a preliminary 3D extinction map 
in the outer disk of the Galaxy. Fig.\,2 shows \ebv~as a function of distance from the sun
and Galactic longitude at $b=-2^\circ$, 0$^\circ$ and $2^\circ$. 
In spite of the limited sky coverage, distinct features, such
as the Perseus and Outer Arms, and effects of warps of the outer disk, are clearly visible.
Given the multi-band reddening of the stars, \rv~values are also estimated and their
spatial variations are investigated in Fig.\,3.
A median value of \rv~=~3.2 is obtained, consistent with previous results.
No obvious spatial variations of \rv~are detected, indicating that dust properties do 
not change significantly in the outer disk.

The LSS-GAC will obtain low-resolution spectra and basic stellar parameters for 
a statistically complete sample of $\gtrsim 3$\,M stars in a large contiguous sky area 
($150 \leq l \leq 210^{\circ}$, $|b| \leq 30^{\circ}$).
For $|b| > 3.5^{\circ}$, the LSS-GAC plans to sample 1,000 stars per sq.deg. 
For $|b| \leq 3.5^{\circ}$, the sampling is doubled.
Combining spectroscopic data from the LAMOST and SDSS, photometric data from 
the GALEX, SDSS, the Xuyi Schmidt Telescope Photometric Survey of the Galactic Anti-center 
(XSTPS-GAC; Liu et al. this volume), Pan-STARRS (Kaiser 2004), 2MASS and WISE,
and spectro-photometric distances and Gaia parallaxes in the future, 
we will produce high spatial resolution (about 10 arcmin), multi-band 
extinction maps in the Galactic anti-center, and then study the distribution of dust and variations of extinction laws.

\begin{figure}[t]
\begin{center}
 \includegraphics[width=5in]{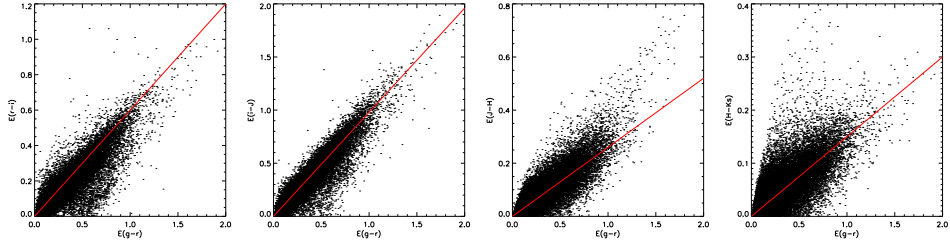} 
 \caption{
Reddening of $r-i$, $i-J$, $J-H$ and $H-Ks$ colors versus that of $g-r$ for 225,422 stars from LSS-GAC
that have XSTPS-GAC, 2MASS photometry
and LAMOST spectral ${\rm S/N}(\lambda4650) > 20$ per pixel.
To avoid crowdness, only one-in-five stars are shown.
For comparison, relations obtained in Yuan, Liu \& Xiang (2013) are over-plotted.
}
   \label{NNN:fig1} 
\end{center}
\end{figure}

\begin{figure}[t]
\begin{center}
 \includegraphics[width=5in]{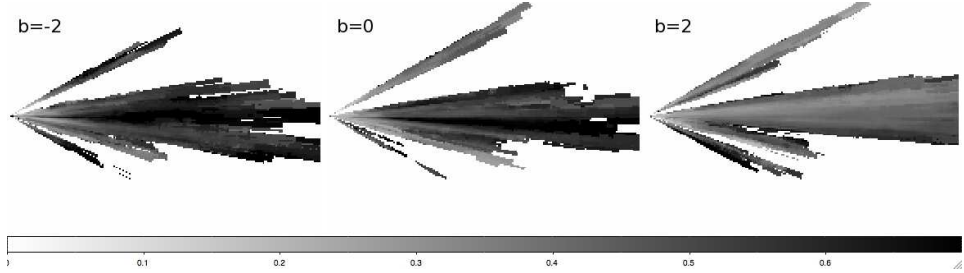} 
 \caption{
\ebv~as a function of distance from the sun (ranging from 0 -- 10 kpc) 
and Galactic longitude (ranging from 150$^\circ$ -- 210$^\circ$) in the Galactic outer disk. 
}
   \label{NNN:fig2} 
\end{center}
\end{figure}

\begin{figure}[t]
\begin{center}
 \includegraphics[width=3.5in]{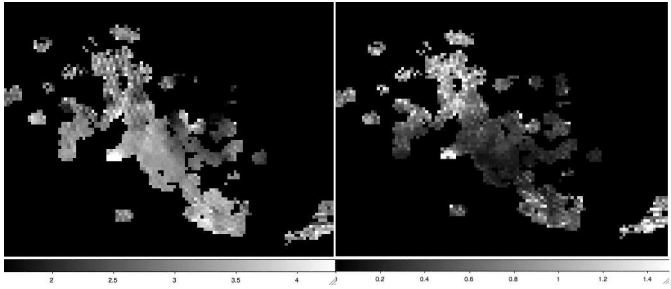} 
 \caption{
Spatial distributions of \rv~(left) and its scattering (right).
The x-axis and y-axis are Galactic longitude ranging from 230$^\circ$ -- 120$^\circ$ and
latitude ranging from $-$40$^\circ$ -- 50$^\circ$, respectively.
}
   \label{NNN:fig3} 
\end{center}
\end{figure}

\section{The diffuse interstellar bands}
DIBs are weak absorption features detected in the
spectra of reddened stars from the near UV to the near IR. DIBs have been
discovered for almost a century, and to date over 400 DIBs have been detected
in Galactic and extragalactic sources (e.g. Hobbs et al. 2008, 2009), 
but none of their carriers is identified (Herbig 1995; Sarre 2006). 
The nature of DIBs remains one of the most challenging problems in
astronomical spectroscopy.
 
Most recent work to identify and investigate the properties and
carriers of DIBs concentrates on high-resolution spectroscopy of a small number
of selected sight-lines. Using a template subtraction method based on the standard pair technique, 
Yuan \& Liu (2012) have successfully identified the DIBs~$\lambda\lambda$5780, 6283 in the
SDSS low-resolution spectra of a sample of about 2,000 stars and
measured their strengths and radial velocities. The sample is by far the
largest ever assembled. The targets span a large range of reddening, 
\ebv~$\sim$~0.2 -- 1.0, and are distributed over a large sky area and involve a wide range
of stellar parameters, confirming that the carriers of DIBs are ubiquitous in the diffuse ISM. 
The sample is used to investigate relations between strengths of
DIBs and magnitudes of line-of-sight extinction, yielding results
(i.e., $EW$(5780) $= 0.61~\times$~\ebv~and $EW$(6283) = 1.26~$\times$~\ebv)
consistent with previous studies (e.g. Friedman et al. 2011). 

DIB features have also been detected in the LAMOST spectra (Fig.\,4) 
of resolving power similar to that of SDSS. In the commissioning spectra of an emission line star, 
even 9 DIBs have been detected (see Fig.\,5 from Yuan \& Liu 2012).
Detections of DIBs towards hundreds of thousands of stars are expected with LAMOST. 
The huge DIB database will provide an unprecedented opportunity to study the demographical 
distribution of DIBs. When combined with other data-sets, it will enable us to address questions like:
How the properties of DIBs (e.g. DIB-DIB, DIB-Extinction, DIB-Gas relations) depend on 
local environment (e.g. UV radiation field, \rv,  extinction in the FUV and the 2175\AA~extinction bump)?
Where are their carriers formed? Can they be formed in the  circumstellar environments?
Meanwhile, DIBs and atomic absorption lines from the ISM can act as good tracers 
to probe the distribution and properties of the ISM and dust.

\begin{figure}[t]
\begin{center}
 \includegraphics[width=4in]{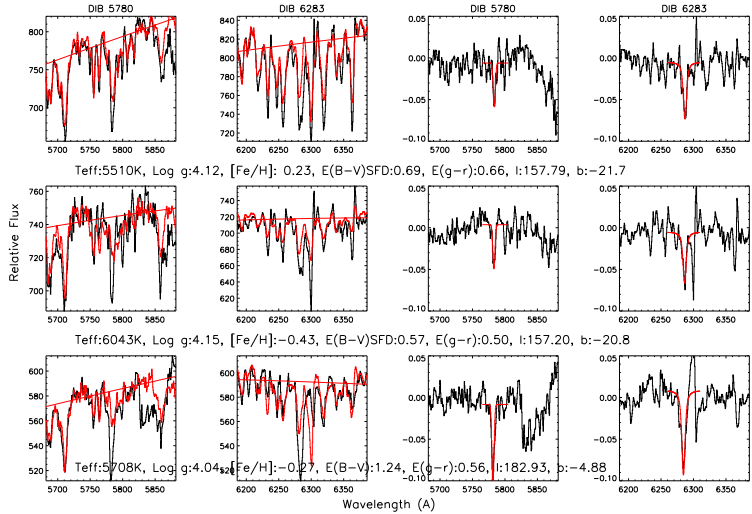} 
 \caption{
The DIBs~$\lambda$$\lambda$5780, 6283 detected in the LAMOST spectra of
three F/G dwarfs. Basic information of each star is listed at the bottom of each row. 
For each row, segments of the target and the scaled template spectra are shown in the left two panels.
The residuals are plotted in the right two panels, with the line fitting results over-plotted.
}
   \label{NNN:fig4} 
\end{center}
\end{figure}

\section{Search for extremely faint emission line nebulae}

Using $\sim$~1.7\,M spectra from the SDSS DR7 (Abazajian et al. 2009), we have undertaken a systematic search for
Galactic planetary nebulae (PNe) via  detections of the \foiii~$\lambda\lambda$4959, 5007 lines (Yuan \& Liu, submitted).
Examples of the SDSS spectra with well detected \foiii~$\lambda\lambda$4959, 5007 lines are shown in Fig.\,5.
Thanks to the excellent sensitivity of the SDSS spectroscopic surveys,
this is by far the deepest search for PNe ever taken, reaching a surface brightness of the \foiii~$\lambda$5007 line S$_{5007}$
down to about 29.0 magnitude arcsec$^{-2}$.
The search recovers 14 previously known PNe in the Galactic Caps.
Most of them are clearly visible on the SDSS broad-band images owe to their high surface brightness.
In total, about 60 new planetary nebula (PN) candidates are identified, including 7 probable
candidates of multiple detections.
All the probable candidates are extremely large (between 21 -- 154\,arcmin) and faint,
located mostly in the low Galactic latitude region with a kinematics similar to disk stars,
confirming the presence of a significant population of previously undetected, large, nearby,
highly evolved PNe in the solar neighborhood.
Four of the candidates have angular sizes between 84 -- 154\,arcmin, and might well be
the largest PNe ever reported.  Based on sky positions and kinematics,
12 of the possible candidates probably belong to the halo population.
If confirmed, they will double the numbers of known PNe in the Galactic halo.
Most newly identified PN candidates are very faint, with S$_{5007}$ between 27.0 -- 30.0 magnitude arcsec$^{-2}$,
and very challenging for previously employed techniques (e.g. slitless spectroscopy, narrow-band imaging).
They greatly increase the number of faint PNe and may well represent the "missing" PN population.

The results have demonstrated the power of large scale fiber spectroscopy in hunting for ultra-faint PNe
and other types of nebulae by detecting nebular emission lines.
The detection limits can be further increased by applying the same method which is used to detect DIBs to the 
SDSS and LAMOST stellar spectra. 
With millions of spectra from the SDSS, LAMOST and other projects,
it will provide a statistically meaningful sample of ultra-faint and large PNe as well as new supernova remnants 
to improve their censuses.

\begin{figure}[t]
\begin{center}
 \includegraphics[width=4in]{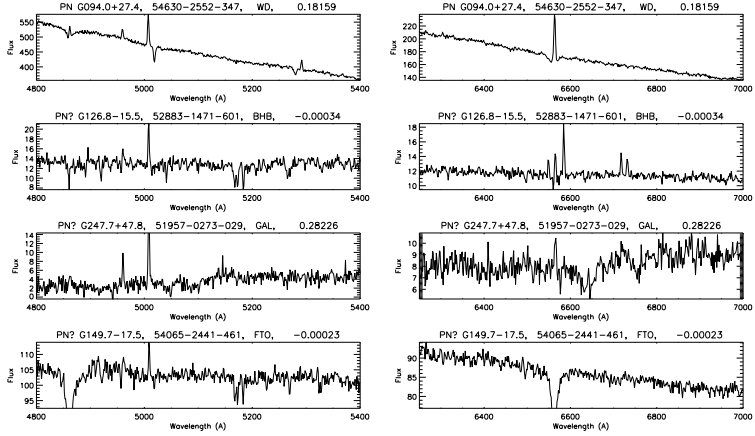} 
 \caption{
Examples of the SDSS spectra with well detected \foiii~$\lambda\lambda$4959, 5007 lines from Galactic PNe and PN candidates.
The PNG identification, SDSS spectral ID,
initial target type and redshift are labeled.
The fluxes are in unit of 10$^{-17}$ergs cm$^{-2}$ s$^{-1}$ \AA$^{-1}$.
}
   \label{NNN:fig5} 
\end{center}
\end{figure}

\section{Photometric calibration of wide field imaging surveys}

Uniform photometric calibration plays a central role in the large-scale
imaging surveys, such as the SDSS, the Dark Energy Survey (DES; The DES Collaboration 2005), Pan-STARS and LSST (Ivezic et al. 2008).
The Stellar Locus Regression (SLR) method (High et al. 2009), adopted in DES, 
can make one wholesale correction for differences in instrumental response, for atmospheric
transparency, for atmospheric extinction, and for Galactic extinction
by adjusting the instrumental broadband colors of stars to bring them
into accord with a universal stellar color-color locus, yielding calibrated colors accurate to a few percent.
The SLR method assumes the standard stellar locus is universal, which is however not always true, 
due to varying stellar populations and extinction, especially in the Galactic disk region.
It also requires a blue filter in addition to at least two of any other filters.

To overcome the limitations above, we propose a new method to perform 
photometric calibration using star pairs from large scale spectroscopic surveys (Yuan et al. in prep), such as LSS-GAC.
The star pairs here are composed of target and control stars from uncalibrated  and calibrated fields, respectively.
This method requires that 1) Extinction values of the targets are known,
which can be from Schlegel et al. (1998) or derived from existing photometric data; 
2) The reddening laws do not change in one field; and 
3) There are a few calibrated fields to obtain the intrinsic colors. 
Then it makes one wholesale correction by adjusting the instrumental colors of target stars to bring them
into accord with their intrinsic colors, and obtain the reddening coefficient simultaneously.
If the accuracy of instrumental colors is about 1\% and about 100 
star pairs can be selected for each field, this method will yield a color calibration accuracy about a few mmag.
It is also useful in checking and improving calibration accuracies of existing surveys.

\begin{discussion}

\discuss{Newberg}{What will be the final spatial resolution of your 3D extinction maps?}

\discuss{Yuan}{About 10 arcmin for a spectroscopic sampling of 1,000 -- 2,000 stars per sq.deg.}
\end{discussion}

\end{document}